\documentclass[10pt,letterpaper,twocolumn]{revtex4} %% two column, final layout

\usepackage{graphicx}
\usepackage[draft]{hyperref}
\usepackage{amsmath}

\begin{document}

%\twocolumn[ %% activate for two-column option
\title{Design of Metamaterial Surfaces with Broad-band Absorbance}

\author{Chihhui Wu and Gennady Shvets}
\affiliation{Department of Physics, The University of Texas at Austin, Austin, TX 78712}

\begin{abstract}
A simple design paradigm for making broad-band ultra-thin plasmonic
absorbers is introduced. The absorber's unit cell is composed of
sub-units of various sizes, resulting in nearly 100\% absorbance
at multiple adjacent frequencies and high absorbance over a broad
frequency range. A simple theoretical model for designing
broad-band absorbers is presented. It uses a single-resonance
model to describe the optical response of each sub-unit and
employs the series circuit model to predict the overall response.
Validity of the circuit model relies on short propagation lengths
of the surface plasmons.
\end{abstract}

%\ocis{240.6680, 310.3915, 310.6628, 310.6805}
% ] %% activate for two-column option

\maketitle %\maketitle must follow title, authors, abstract and \pacs

Plasmonic metamaterials (MMs) are subwavelength metallic
nanostructures designed to provide extraordinary electromagnetic
properties in specified frequency
ranges~\cite{Lindell_2001,schurig_prl2003}. The combination of the
plasmonic response of conduction electrons and appropriate (but
not necessarily geometrically complicated) design of the
structures results in subwavelength feature sizes and strong
near-field enhancement. These properties pave the way to various
applications, such as surface enhanced Raman
scattering~\cite{Fei_2008,Paul_2008}, light
trapping~\cite{Atwater_NM2010}, bio-molecule
sensing~\cite{adato_pnas09}, and plasmonic
absorbers~\cite{padilla_prl08,wu_spie08,avitzour_prb09,giessen_nl10,padilla_apl10,wu_2011}.
For the reasons going beyond the simplicity of fabrication, many
applications utilize MM surfaces instead of MMs. For example, MM
surfaces provide solid/liquid interfaces for bio-sensing, or can
be used as ultra-thin coatings to provide spectrally-selective
optical properties (e.g, absorption).

\begin{figure}[h]
 \centering
 \includegraphics[width=.45\textwidth]{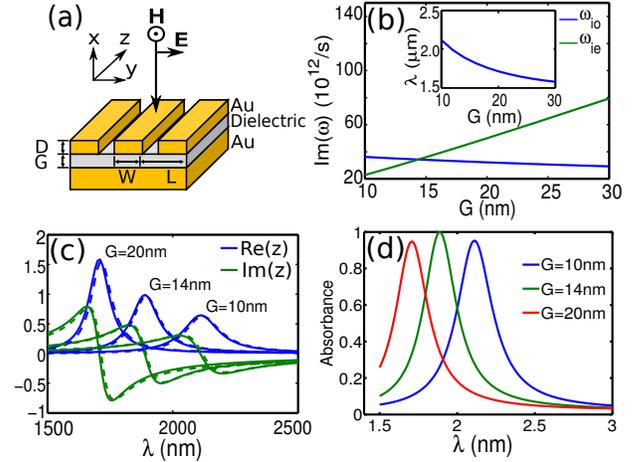} \
 \caption{(a) Schematic of the Narrow Band Metamaterial Absorber
 (NBMA). (b) Dependence of radiative ($\omega_{ie}$) and resistive
 ($\omega_{io}$) decay rates on $G$. Remaining dimensions:
 $L=350$~nm, $W=250$~nm, and $D=20$~nm. Inset: resonant wavelength
 vs $G$. (c) Impedances of NBMAs for $3$ values of $G$. Solid lines:
 from driven simulations, dashed lines: from the single oscillator
 model. (d) Absorbance of the NBMAs with same values of $G$ as in
 (c).}\label{fig:1}
\end{figure}

Simple designs of narrow-band MM absorbers (NBMA) comprised of an
array of plasmonic strips/patches separated by a thin dielectric
spacer from a ground plate, as illustrated in
Fig.~\ref{fig:1}(a), have been recently proposed and experimentally
implemented~\cite{wu_spie08,giessen_nl10,padilla_apl10,wu_2011}. 
``Perfect'' ($100\%$ at the peak) light absorption by
NBMAs is enabled by the resonant excitation of a surface plasmon
polariton (SPP) whose frequency is controlled by the structural dimensions. Using such
NBMAs as building blocks for a broad-band metamaterial absorber
(BBMA) appears straightforward: combining several such blocks
($i=1,...N$) with spectrally-close absorption peaks at
$\omega_0^{(i)}$ and bandwidths $\Delta \omega_0^{(i)}$ into a
macro-cell should deliver the desired broad-band feature as long
as $|\omega_0^{(i+1)} - \omega_0^{(i)}| < \Delta \omega_0^{(i)}$.
In reality, designing BBMAs using this
approach~\cite{padilla_prl11} has been challenging. For example,
the individual absorption peaks of the constitutive NBMAs can each
yield ``perfect'' absorption, but at the expense of large spectral
separation. Alternatively, constitutive NBMA's with
spectrally-close absorption peaks can fail to provide ``perfect''
absorption at every peak. Systematic approach to designing NBMAs
and combining them into a unit cell of a BBMA is needed. In this
Letter, we provide such an approach based on a simple impedance
model. We show that the surface impedance $z$ of a BBMA can be
obtained from the impedances $z_i$ of the constitutive NBMAs using
a simple additive formula (see Eq.~(\ref{eq:series})), which is
valid as long as the SPPs' propagation lengths are much smaller
than the separation between NMBAs.

We start by introducing a simple single-resonance model of an NMBA
based on the periodic structure shown in Fig.~\ref{fig:1}(a). The
structure coupled to the incident radiation is described within
the framework of a single resonator model~\cite{haus_1984}. An
open resonator is defined by its natural frequency, $\omega_0$,
and finite lifetimes, $\tau_o=1/\omega_{io}$ and
$\tau_e=1/\omega_{ie}$, respectively determined by the Ohmic and
radiative losses. The interaction between the resonator and the
incident field of amplitude $E_I$ is described by the following
equations:
\begin{eqnarray}
&&\frac{d}{dt}a = -i\omega_0~a - (\omega_{io}+\omega_{ie})~a +
\sqrt{2~\omega_{ie}}~E_I, \nonumber
\\ &&E_R = -E_I + \sqrt{2~\omega_{ie}}~a, \label{eq:oscillator}
\end{eqnarray}
where $a$ and $E_R$ are the amplitudes of the resonator and the
reflected field, respectively. The specific geometry of the
resonator that includes the ground plate is manifested in the
expression for $E_R$, which is a superposition of the resonator and
ground plate reflection. Within this simple single-resonance
model, the surface impedance defined as $z \equiv (1+r)/(1-r)$,
where $r=E_R/E_I$, is given by:
\begin{eqnarray}
&&z = \frac{\omega_{ie}}{-i(\omega - \omega_0) + \omega_{io}},
\label{eq:impedance}
\end{eqnarray}
and all optical properties of the NBMA are determined by three
parameters: $\omega_0$, $\omega_{ie}$, and $\omega_{io}$. These
parameters depend on the structural dimensions and can be obtained
by calculating the ``leaky'' SPP eigenmodes~\cite{wu_2011} using
eigenvalue simulations (i.e., no incident field). By tuning the
geometric parameters of the NBMA, one can engineer $\omega_0$,
$\omega_{ie}$, and $\omega_{io}$ and achieve $z=1$ (and,
therefore,perfect absorption) whenever the critical coupling
condition $\omega_{ie}=\omega_{io}$ is satisfied.

The validity of the single-resonance model was verified for
several NBMAs as shown in Fig.~\ref{fig:1}. Plasmonic strips and
ground plate are assumed to be made of gold (described by the
Drude permittivity $\epsilon=1 - \omega_p^2/\left[\omega(\omega +
i\omega_c)\right]$, where $\omega_p=1.32\times 10^{16} rad/s$ and
$\omega_c=1.2\times 10^{14} rad/s$) and the spacer is assumed to
have a non-dispersive dielectric permittivity $\epsilon_d = 2.25$.
The accuracy of the $z(\omega)$ obtained from
Eq.~(\ref{eq:impedance}) was verified by comparing it to the
impedance obtained from the reflection coefficients calculated at
various frequencies $\omega$ of the incident radiation. Excellent
agreement shown in Fig.~\ref{fig:1}(c) validates the
single-frequency model which is highly computationally efficient
because only a single eigenvalue/eigenmode simulation is
required~\cite{wu_2011} to determine $z(\omega)$ for all
frequencies.

The tunability of the NBMA's frequency characteristics
$(\omega_0,\omega_{ie},\omega_{io})$ with respect to the spacer
thickness $G$ is illustrated in Fig.~\ref{fig:1}(b). Note from
Fig.~\ref{fig:1}(d) that considerable peak absorbances (in excess
of $90\%$) is achieved for a wide range of spacer thicknesses
$10{\rm nm} < G < 20 {\rm nm}$. The smooth dependence of the
frequency characteristics on $G$ shown in Fig.~\ref{fig:1}(b) (as
well as on the other structure parameters) enables rapid design of
the appropriate parameter set ($W$, $L$, $G$, $D$) of the NBMA
that provides the targeted $(\omega_0,\omega_{ie},\omega_{io})$.

\begin{figure}[h]
 \centering
 \includegraphics[width=.48\textwidth]{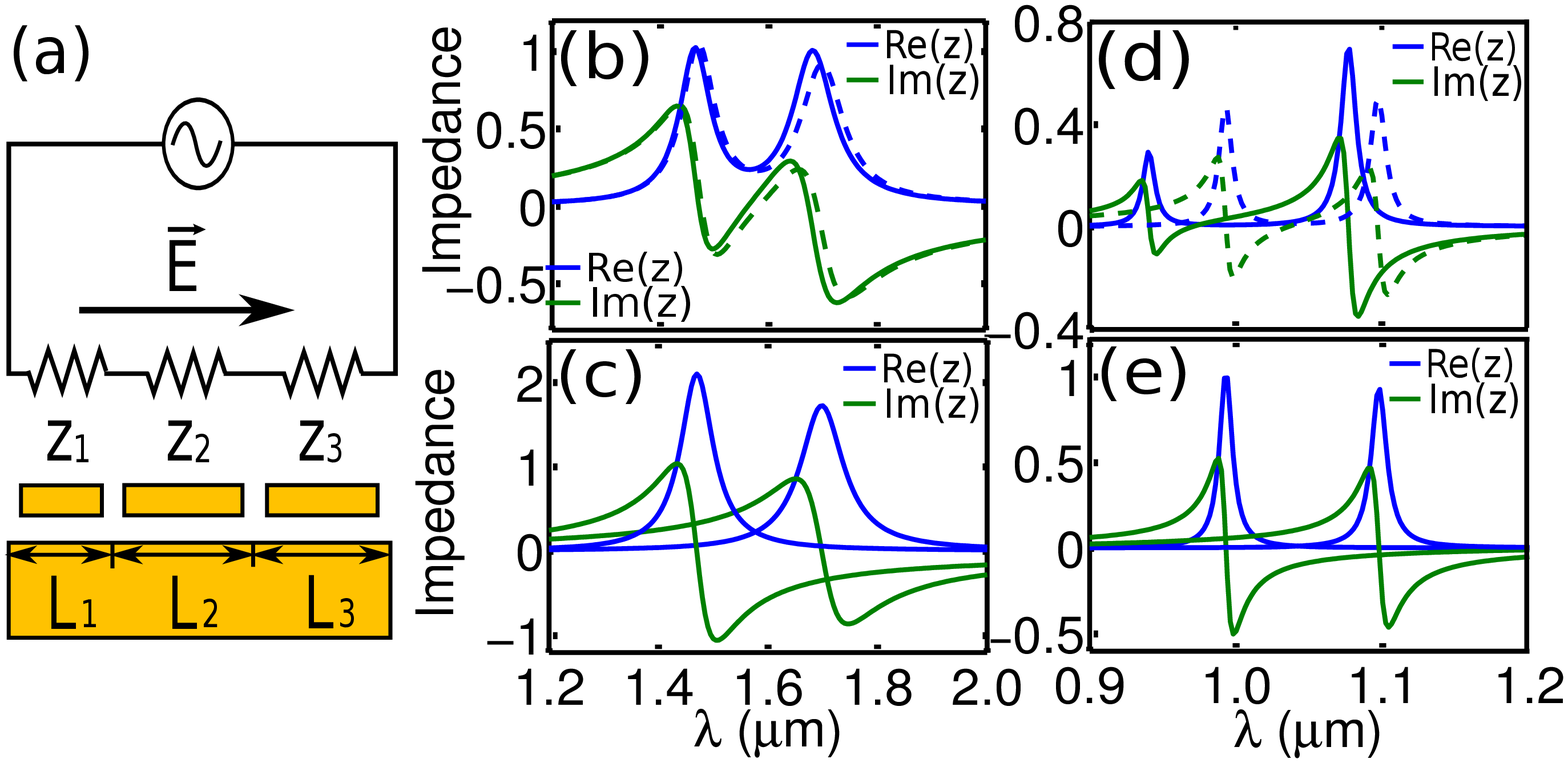} \
 \caption{(a) Illustration of the circuit model used to derive
 Eq.~(\ref{eq:series}). (b,d) Impedance $z$ of the resonant
 surface with a macro-cell comprised of two sub-units.
 Sub-unit
 dimensions: (b) [$L_1$, $L_2$, $W_1$, $W_2$] = [407nm, 383nm,
 268nm, 227nm] and [D,G]=[17nm, 28nm]; (d) [$L_1$, $L_2$, $W_1$,
 $W_2$] = [432nm, 371nm, 409nm, 356nm] and [D,G]=[13nm, 15nm].
 Solid lines in (b,d): $z$ from direct simulations; dashed lines:
 $z$ from Eq.~(\ref{eq:series}), with individual impedances
 $z_{1,2}$ plotted in (c,e), respectively.}\label{fig:2}
\end{figure}

Two important features of the NBMAs described in Fig.~\ref{fig:1}
make them promising building blocks for designing a BBMA. First,
their period is strongly sub-wavelength: $L_i \ll \lambda_0^{(i)}
\equiv 2\pi c/\omega_0^{(i)}$. That implies that a sub-wavelength
($L = \sum_i L_i \leq \lambda$) unit cell of a BBMA can
accommodate a large number of NBMA-based sub-units satisfying $L_i
\ll L \leq \lambda$. Second, the propagation length $l_{\rm
SPP}^{(i)}$ of the SPP responsible for absorption satisfies
$l_{\rm SPP}^{(i)} < L^{(i)}$. This property has been
shown~\cite{wu_2011} to be related to the wide-angle absorbance of
the NBMA. While these two conditions are related, they are not
equivalent as shown below. Nevertheless, when both conditions are
satisfied, a simple expression for a the surface impedance of a
BBMA with a macro-cell composed of $N$ NBMA's can be obtained:
\begin{eqnarray}
&&z=\sum\limits_{i=1}^N z_i L_i/\sum\limits_i L_i,
\label{eq:series}
\end{eqnarray}
where $z_i$ ($L_i$) is the impedance (period) of each periodic
constitutive NBMA. Derivation of Eq.~(\ref{eq:series}), based on
the series combination model~\cite{engheta_07}, is illustrated in
Fig.~\ref{fig:2}(a), where the external field represents a voltage
drop and each NBMA resonator is treated as a circuit element. The
key assumption is that each component of the circuit
functions independently from each other, i.e., there is little
mutual capacitive or inductive coupling between adjacent
sub-units. This condition is satisfied owing to the second feature
of NBMAs. Below we verify Eq.~(\ref{eq:series}) for two cases: (i)
strongly (Fig.~\ref{fig:2}(b,c)) and (ii) weakly
(Fig.~\ref{fig:2}(d,e)) confined SPPs of the individual sub-units.
In each case the unit cell of a candidate BBMA consists of two
sub-units with dimensions stated in Fig.~\ref{fig:2}.

In case (i), excellent agreement between BBMA's impedance
calculated from Eq.~(\ref{eq:series}) and from direct numerical
simulations is shown in Fig.~\ref{fig:2}(b). Each sub-unit,
representing a resonant high-impedance ($z_{1,2} \approx 2$)
surface, is independent of the other inside the macro-cell because
$l_{\rm SPP} \approx 360 {\rm nm} < L_{1,2}$. The impedance of the
combined candidate BBMA has two spectral peaks with values $z
\approx 1$ corresponding to $100\%$ absorption. On the other hand,
Eq.~(\ref{eq:series}) clearly fails in case (ii) as shown in
Fig.~\ref{fig:2}(d). That is because $l_{\rm SPP} \approx 705 {\rm
nm} > L_i$, and the two sub-units strongly interact inside the
macro-cell, thereby invalidating the series impedance model. Note
that to achieve such long propagation length of the SPP, the
electron collision frequency, $\omega_c$, in Au has been
artificially reduced by a factor of four.

\begin{figure}[h]
 \centering
 \includegraphics[width=.48\textwidth]{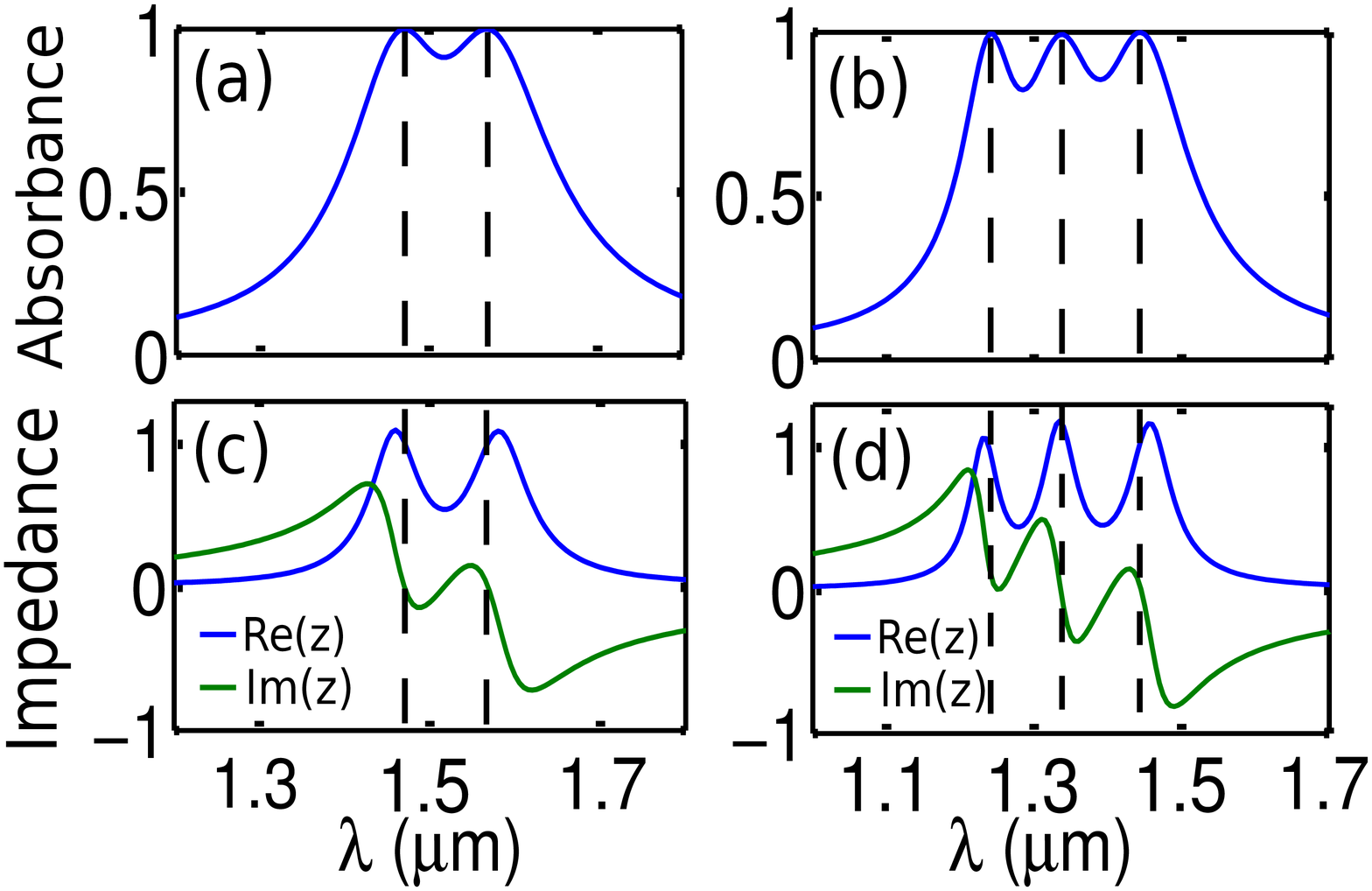} \
 \caption{Absorption spectrum (top row) and surface impedance
 (bottom row) of the designed Broad Band Metamaterial Absorbers
 (BBMA) with two (left
 column) and three (right column) sub-units per unit cell.
 Sub-units dimensions:
 [$L_1$, $L_2$, $W_1$, $W_2$] = [448nm, 348nm, 249nm, 226nm] in
 (a,c); [$L_1$, $L_2$, $L_3$]=[327nm, 280nm, 271nm] and [$W_1$,
 $W_2$, $W_3$] = [224nm, 200nm, 180nm] in (b,d). Metal and spacer
 thickness [D, G]=[17nm, 28nm] is the same for all
 sub-units.}\label{fig:3}
\end{figure}

We now present two examples of BBMAs constructed by engineering
their unit cells from several (two in Fig.~\ref{fig:3}(a,c) and
three in Fig.~\ref{fig:3}(b,d)) NBMA sub-units. We use the example
of two sub-units inside the unit cell of a BBMA to describe the
computational procedure for obtaining an absorber with two $100\%$
absorption peaks at the prescribed frequencies $\omega_1$ and
$\omega_2$. That corresponds to $z(\omega_1) = z(\omega_2) =1$,
where $z(\omega)$ is calculated from Eq.~(\ref{eq:series}). The
resulting four equations (note that $z$ is a complex number) must
be satisfied for four unknown. For simplicity, we assume $D_1=D_2$
and $G_1=G_2$ to be fixed. This leaves the complex impedance
$z=z(L_1,L_2,W_1,W_2)$ as a function of four unknowns that can be
numerically solved for to satisfy the above four equations. The
numerical implementation based on the multi-dimensional Newton's
method is greatly simplified by the variables' separation in
Eq.~(\ref{eq:series}): the number of partial derivatives $\partial
z/\partial L_i$,  $\partial z/\partial W_i$ is reduced because
$\partial z_i/\partial (L,W)_j =\delta_{i,j} \partial z_0/\partial (L,W)$, 
where $z_0$ is the impedance of one single sub-unit.

The results of such numerical calculations for a BBMA with two
sub-unit per unit cell are shown in Fig.~\ref{fig:3} (left column),
where the resulting double-peaked impedance $z$ is plotted in
Fig.~\ref{fig:3}(c) and the corresponding frequency-broadened
absorptivity is plotted in Fig.~\ref{fig:3}(a). A similar
numerical procedure was employed to design a triple-peaked BBMA
illustrated in Fig.~\ref{fig:3} (right column; all sub-unit
parameters in the caption). Note that over $90\%$ absorptivity is
achieved over the spectral range of $1.2\mu m < \lambda < 1.5\mu
m$. Note that in both examples presented in Fig.~\ref{fig:3} the
periodicity $L_{\rm BBMA}=\sum_i L_i$ of the BBMA is
sub-wavelength: $L_{\rm BBMA}< \lambda$. It is essential to
satisfy this condition to avoid the emergence of the additional
diffraction orders which create additional radiative loss channels
and further complicate the impedance model. In practice, this sets
a limit on the attainable bandwidth. For example, for the
strip-based designs presented in this Letter, the number of
sub-units is limited to $3$ (or $9$ for a patche-based design). A
possible solution could involve an aperiodic distribution of
sub-units. Based on negligible interaction between the sub-units,
it is expected that Eq.~(\ref{eq:series}) should hold.

In conclusion, we have demonstrated that a broad-band metamaterial
absorber (BBMA) can be designed by constructing a super-lattice
structure with each sub-lattice absorbing at different
frequencies. Based on a single oscillator model and a series
circuit model, broad-band absorbers with multiple perfectly
absorbing peaks are shown to be attainable, and a procedure for
designing such absorbers is provided. Examples of BBMAs comprised
of two and three sub-units operating in the telecommunications
frequency range are presented, and an average $90\%$ absorbance in
the $1.2\mu m < \lambda < 1.5\mu m$ range is predicted. Such
ultra-thin ($<100$nm) BBMAs can be used for infrared detection and
other applications. This work is supported by the Office of Naval
Research (ONR) grant N00014-10-1-0929.

\end{document}